# 3-D Coherent Multi-Transducer Ultrasound Imaging with Sparse Spiral Arrays

Laura Peralta, *Member, IEEE*, Daniele Mazierli, *Student Member, IEEE,* Alberto Gomez, Joseph V. Hajnal, Piero Tortoli, *Fellow, IEEE,* and Alessandro Ramalli, *Senior Member, IEEE*

*Abstract—* Coherent multi-transducer ultrasound (CoMTUS) creates an extended effective aperture through the coherent combination of multiple arrays, which results in images with enhanced resolution, extended field-of-view, and higher sensitivity. The subwavelength localization accuracy of the multiple transducers required to coherently beamform the data is achieved by using the echoes backscattered from targeted points. In this study, CoMTUS is implemented and demonstrated for the first time in 3-D imaging using a pair of 256-element 2-D sparse spiral arrays, which keep the channel-count low and limit the amount of data to be processed. The imaging performance of the method was investigated using both simulations and phantom tests. The feasibility of free-hand operation is also experimentally demonstrated. Results show that, in comparison to a single dense array system using the same total number of active elements, the proposed CoMTUS system improves spatial resolution (up to 10 times) in the direction where both arrays are aligned, contrast-to-noise-ratio (CNR, up to 30%), and generalized CNR (up to 11%). Overall, CoMTUS shows narrower main lobe and higher contrast-to-noise-ratio, which results in an increased dynamic range and better target detectability.

## I. INTRODUCTION

MEDICAL ultrasound (US) is a low-cost imaging method that is long-established and widely used for screening, diagnosis, therapy monitoring, and guidance of interventional procedures. However, the usefulness of conventional US systems is limited by physical constraints that lead to low-resolution images with a restricted field of view (FOV) and view-dependent artefacts. All these limitations, which are especially prevalent when imaging at depth and are more pronounced in an increasingly obese population [1], stem from the limited extent of the transmitting and receiving apertures [2]. Indeed, the image resolution of US is diffraction limited and mostly depends on three factors: the relative aperture size, the central frequency of the transmitted wave, and the imaging depth [3]. Although high frequency waves can achieve better resolution, they are more easily absorbed and attenuated by tissues. Therefore the use of low frequencies is the only choice when imaging deep in the body. Consequently, a large aperture size is required to improve resolution, penetration, and FOV.

To enlarge the FOV, multiple images can be acquired either by the same probe [4][5] or by different probes [6] and incoherently compounded using image registration. These image fusion techniques based on the exploitation of multiple points of view can reduce view-dependent artefacts [7] and improve contrast [5]. Recent studies have demonstrated that the contrast improvements also enhance motion and strain estimations from US image data [8][9]. However, such approaches do not improve the diffraction-limited resolution and they cannot directly achieve the gains in sensitivity that could be realized if the multiple arrays were coherently combined into a one large effective aperture [10].

Recently, we have developed coherent-multi transducer ultrasound (CoMTUS) imaging [11], which coherently combines the radio frequency (RF) data received by multiple synchronized transducers that take turns transmitting plane waves (PWs) into a common FOV. In contrast to previous proposed techniques, which rely on external tracking devices [10] or on fixed and known geometries [8], [9], CoMTUS achieves subwavelength localization accuracy, required for coherent compounding, by directly using the backscattered sound field of multiple transducers. CoMTUS estimates the optimal beamforming parameters, which include the relative transducer locations and the local speed of sound, by maximizing the coherence of the received echoes, resulting from different targeted scatterers in the medium, by cross-correlation. This allows the operation of a large aperture with some tolerance for acoustically heterogeneous tissues [12]. The approach has been previously experimentally validated in 2-D imaging, using two linear arrays that are constrained to the same elevational plane. However, the alignment of the arrays within

This research was funded by the Royal Society (URF/R1/211049) and Wellcome Trust/EPSRC iFIND project IEH Award (102431) (www.iFINDproject.com). The authors acknowledge financial support from the Department of Health via the National Institute for Health Research (NIHR) comprehensive Biomedical Research Centre award to Guy's & St Thomas' NHS Foundation Trust in partnership with King's College London and Kings College Hospital NHS Foundation Trust. *(Corresponding author: Laura Peralta).*

Laura Peralta is with the Department of Surgical & Interventional Engineering, School of Biomedical Engineering & Imaging Sciences, King's College London, St Thomas' Hospital, London SE1 7EH, U.K. (e-mail: laura.peralta_pereira.@.kcl.ac.uk).

Daniele Mazierli, Piero Tortoli and Alessandro Ramalli are with the Department of Information, Università degli Studi di Firenze, 50121 Florence, Italy.

Joseph V. Hajnal and Alberto Gomez are with the Department of Biomedical Engineering, School of Biomedical Engineering & Imaging Sciences, King's College London, St Thomas' Hospital, London SE1 7EH, U.K.

the same imaging plane presents a challenge for in vivo applications [13]. In free-hand scanning, the alignment of the probes with the required precision is not possible and the use of a rigid holder limits the flexibility and operability of the system. Extending CoMTUS to 3-D imaging may mitigate misalignment issues. In addition, the generation of volumetric images can provide more complete information and higher diagnostic accuracy than 2-D imaging [14][15].

Real-time volumetric ultrasound became possible with the development of 2-D matrix-arrays, which can steer and scan the beam in two dimensions and achieve 3-D volumetric images with good temporal resolution [16]. However, 2-D matrix arrays usually require thousands of elements which must be individually controlled, and this results in complex systems that are demanding of hardware electronics [17]. In addition, a significantly larger amount of received data than in 2-D imaging systems need to be stored and processed. Moreover, specifically for CoMTUS, which relies on multiple transducers, the required number of active elements can become prohibitive.

To address the above challenges in 3-D ultrasound systems, different methods have been proposed such as microbeamforming techniques [18], multiplex matrix array probes [19], row-column-addressed arrays [20][21], and sparse arrays [22]. The latter approach uses a reduced number of elements whose positions are selected following random [23], optimized [24], or deterministic distributions [25][26]. In [26] the elements are aperiodically distributed according to the position of the seeds of a Fermat's spiral array, whose density is modulated according to a tapering function. The feasibility of spiral arrays for 3-D US imaging has been demonstrated in different applications, such as high frame rate echocardiography [27], US super resolution [28], and US Doppler [29][30].

In this study, CoMTUS is implemented and demonstrated for the first time in 3-D imaging using a pair of 2-D spiral arrays with a total of 512 active elements, which are independently controllable by two ultrasound advanced open platform (ULA-OP 256) scanners [31].

The article is organized as follows: Section II presents the design of the sparse arrays. Then, the simulations and experimental methods are described along with the metrics used to evaluate the image quality. Section III shows simulation and experimental results that are finally discussed in Section IV. Conclusions are summarized in Section V.

## II. MATERIALS AND METHODS

### A. Array layouts

Two prototype probes, P1 and P2, were designed based on the $32 \times 35$ gridded layout of a 2-D 1024-element matrix array (Vermon S.A., Tours, France), where the rows 9, 18 and 27 in the y-direction were not connected due to routing requirements. The array had a central frequency of 3.7 MHz and 60% bandwidth with element pitch of 300 μm × 300 μm. For each probe, 512 elements were hardwired as proposed in [27], corresponding to the quasi dense configuration reported in Fig.1a. Since it was not possible to connect all 512 elements to a single ultrasound probe adapter, the 512 elements were divided into two equal sized sub-groups, each connected to an independent connector (labelled A and B). The selection of the 256 elements on connector A was based on an ungridded, 10.4-mm-wide spiral with 256 seeds, whose density tapering was modulated according to a 50%-Tukey window [26] (see Fig. 1b). The remaining 256 elements were linked to connector B.

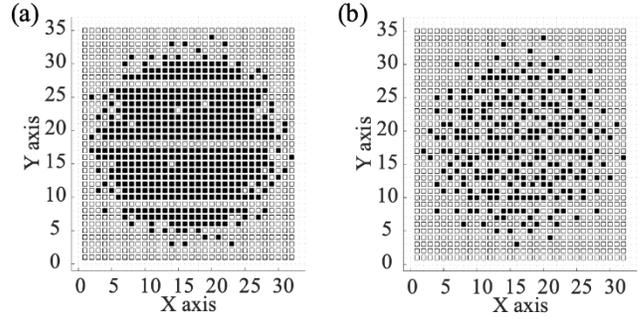

**Fig. 1.** Layout of the 2-D (a) dense and (b) sparse spiral arrays with solid squares showing the hardwired elements. Inactive rows (9, 18, and 27) are due to probe routing requirements.

These element selections, allowing the independent control of 512 elements in both transmission (Tx) and reception (Rx), were ideal for an experimental demonstration of CoMTUS by exploiting two 256-channel systems. In CoMTUS mode, the two systems were connected to connectors A of probes P1 and P2, respectively (layout in Fig.1b). Hereinafter this configuration will be referred as to 2×256Tx/Rx. To provide a reference mode, hereinafter referred to as 1×512Tx/Rx, the two scanners were associated with the connectors A and B of probe P1 (layout in Fig.1a) [32].

### B. Probe configurations and spatial positions

Fig.2 shows the spatial configuration of both sparse arrays in the 2×256Tx/Rx configuration. In this configuration, three different coordinate systems can be identified: a local coordinate system $\{x_1, y_1, z_1\}$ with origin at the center of array P1; a local coordinate system $\{x_2, y_2, z_2\}$ with origin at the center of array P2; and an effective coordinate system, $\{x_{\text{eff}}, y_{\text{eff}}, z_{\text{eff}}\}$, which is defined at the center of the resulting effective aperture created by both arrays.

For both experiments and simulation, the two arrays were aligned on the x-axis to scan a common volume of interest approximately at 40 mm depth. The position of array 2 was set relative to array 1 by using the following geometric transformation,

$$F = \begin{bmatrix} R_{3\times 3} & T_{3\times 1} \\ 0_{1\times 3} & 1 \end{bmatrix} \quad (1)$$

where $T = [34,\ 0,\ 14.4]^T$ mm is the offset between the centers of the arrays, and $R$ is a 3×3 matrix parameterized through three rotation angles, $\{\theta_x, \theta_y, \theta_z\} = \{0º, -50º, 0º\}$.

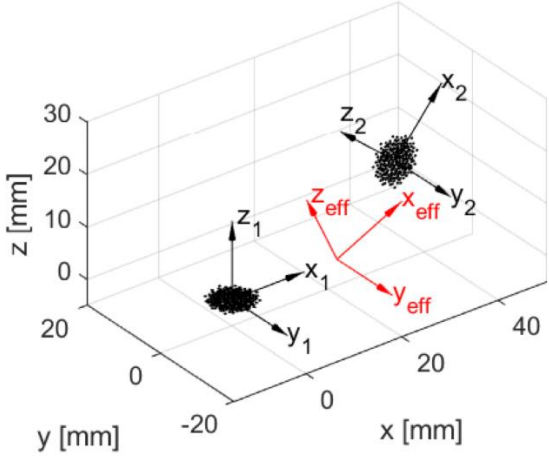

**Fig. 2.** Spatial configuration of the sparse spiral arrays. The different coordinate systems are shown: $\{x_1, y_1, z_1\}$ local coordinate system of array P1, $\{x_2, y_2, z_2\}$ local coordinate system of array P2, and effective coordinate system, $\{x_{\text{eff}}, y_{\text{eff}}, z_{\text{eff}}\}$.

In the 1×512Tx/Rx configuration, only array P1 was used and the coordinate system was placed at the center of the array P1 as usual.

*C. Scan sequences*

The scan sequences were based on the transmission of steered PWs, obtained with four-cycle Gaussian pulses at 3-MHz center frequency and without any Tx/Rx apodization. PWs were steered with a maximum transmit angle of 5°, chosen according to the imaging depth and spatial configuration of the probes [33]. Two different Tx sequences, named as xPW and sPW, were investigated:

1) xPW: 9 PWs were steered along the lateral (*x*-axis) and then the elevational (*y*-axis) directions, with steps of 2.5°, to form a cross-shape over the steering angle domain (see Fig. 3a).
2) sPW: 22 PWs were simultaneously steered in both lateral and elevational directions following, in the steering angle domain, a 2-D spiral distribution to maximize image quality (see Fig. 3b) [34].

In the 2×256Tx/Rx configuration, the scan sequences were implemented in an alternating scheme in which each array in turn transmitted a PW, with the backscattered echoes always simultaneously received by both arrays. In the 1×512Tx/Rx configuration, the same Tx sequences were used, but all the 512 elements were simultaneously used for each pulse repetition interval.

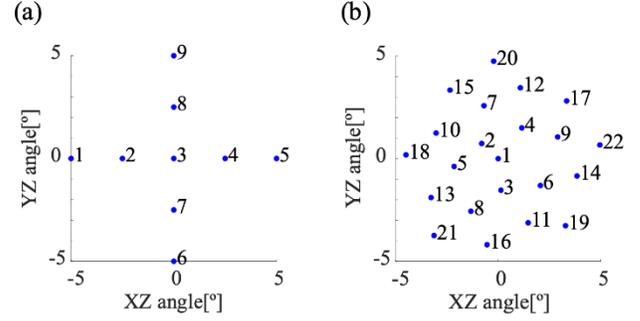

**Fig. 3.** Tx angles adopted in the Tx sequences. (a) 9 PWs steered from -5° to 5°, 2.5° step, first in the lateral (*x*-axis) and then elevational (*y*-axis) directions. (b) 22 PWs steered from -5° to 5° following a spiral distribution. The numbers label the transmit order of the different angles.

*D. Experiments*

The experimental setup consisted of two synchronous 256-channel Ultrasound Advanced Open Platform (ULA-OP 256) scanners (MSD Laboratory, University of Florence, Florence, Italy) [35][36] and two prototype Vermon probes, which were hardwired as described in section IIA. The two ULA-OP 256 were configured so they could implement both the 1×512Tx/Rx and the 2×256Tx/Rx configurations and both the xPW and sPW scan sequences. The pulse repetition frequency (PRF) was set to 3000 Hz and raw channel data were acquired at a sampling frequency of 26 MHz. To facilitate the manual positioning of the probes during the experiments, for each probe, two B-mode images were shown in real-time. They were reconstructed over the two planes perpendicular to the axis of each probe for the xPW sequence.

The two arrays were mounted on a 3-D printed holder that allowed the probes to be kept in an approximately fixed relative configuration as described in section II.B and scanning a common volume of interest of the phantom.

A custom-made phantom was used to experimentally validate the method and characterize resolution and contrast. First, in a beaker, 800g of deionized water were mixed with 100g of glycerin and 28g of agar until dissolved. The mixture was heated up to 90° (boiling point of agar is 85°) while being stirred using a hot plate magnetic stirrer. The solution was then poured into a rectangular mold of dimensions 12 cm × 9 cm × 9 cm. Three steel spheres (300 μm diameter) were carefully placed inside the agar solution to create three hyperechoic point-like regions. After room temperature was reached, the solution filled mold was placed in a fridge and allowed to settle for at least 12 hours. Then, the sample was carefully removed from the mold, keeping the metallic spheres embedded. In a final step, a cylindrical inclusion to mimic an anechoic lesion approximately 6.5 mm in diameter was manually drilled.

During the experiments, the phantom was placed in a water tank at room temperature and positioned so that all spheres and the anechoic region were in the common FOV of the two arrays. Without moving the probes, data were acquired first for the 2×256Tx/Rx configuration and then for the 1×512Tx/Rx one.

*E. Simulations*

The configurations, scan sequences, and array positions described in the previous sections were simulated in MATLAB (The MathWorks, Natick, MA, USA) using Field II [37][38]. The sampling frequency was set to 50 MHz and the speed of sound was set at 1496 m/s.

A numerical phantom was specifically developed to correspond to the experiments and assess imaging performance. It consisted of a background of randomly generated scatterers with a density distribution of 23 scatterers mm³ around an anechoic cylindrical cyst of 6.5 mm diameter, and 3 spherical hyperechoic (echogenicity 65 dB over the background) targets of 300 µm diameter. The targets and the inclusion were placed to approximately match the characteristics of the phantom used in the experiments. The phantom size was 24 mm (*x*-axis) × 10.5 mm (*y*-axis) × 40 mm (*z*-axis) and was centered at y = 4 mm and z = 40 mm in the coordinate system defined by the array P1. The two arrays were placed as described in Section II-B.

*F. Outcome parameters*

To evaluate the system performance CoMTUS images, acquired with the 2×256Tx/Rx configuration were compared with images obtained with the single probe 1×512Tx/Rx configuration.

The backscattered echoes from multiple Tx and Rx positions were first aligned by optimizing their spatial coherence, as described in detail in [11]. CoMTUS images were then beamformed compounding the RF data acquired by both arrays. The images were reconstructed in the effective aperture coordinate system (Fig.2). This coordinate system leads to a more conventional point-spread-function (PSF) shape where the principal resolutions are aligned with the Cartesian axes [11]. Data acquired by the single array, 1×512Tx/Rx configuration, were beamformed in the local coordinate system defined at the center of the probe P1.

Imaging resolution and contrast were assessed from the PSF of the hyperechoic targets and the cylindrical anechoic inclusion. The full width half maximum (FWHM) measured from PSF of the deepest target (located at 46 mm depth in the coordinate system P1), was used to evaluate the resolution in all three cartesian directions. Contrast ratio (CR), contrast-to-noise-ratio (CNR), and generalized contrast-to-noise-ratio (gCNR) [39] were quantified from the cylindrical inclusion and the speckle background as,

$$CR = 20\log_{10}\frac{\mu_i}{\mu_o} \quad (2)$$

$$CNR = \frac{\mu_i - \mu_o}{\sqrt{\sigma_i^2 + \sigma_o^2}} \quad (3)$$

$$gCNR = 1 - OVL \quad (4)$$

where $\mu_i$ and $\mu_o$ are the means of the signal inside and outside of the anechoic region, respectively, and $\sigma_i$ and $\sigma_o$ represent the standard deviations of the signals inside and outside of the anechoic region, respectively. *OVL* is the overlap area between the probability density functions of both regions of signal inside and outside of the anechoic inclusion [39]. As shown in Fig. 4, two cubic regions of 3.5×3.5×3.5 mm³ placed at the center of the inclusion (red squares) and 6.2 mm above (yellow squares), were used for the calculations.

### III. RESULTS

Fig. 4 shows the simulated B-mode images obtained with both the configurations for the sPW scan sequence. The image quality metrics for the simulations are summarized in Table I for both scan sequences. Compared with the 1×512Tx/Rx configuration, overall, the CoMTUS images have a larger FOV, higher contrast metrics, and significantly smaller FWHM in the lateral direction (*x*-axis), which results in an increased dynamic range and better target detectability. In the *xz*-plane, the speckle texture is thinner due to the improvement in lateral resolution. Compared with the xPW sequence, the sPW scan sequence improves all contrast metrics in both configurations. For both sequences, CoMTUS outperforms the single array system in CR, -16.7dB vs -13.8dB (xPW) and -22.1dB vs -17.1dB (sPW), CNR, 1.23 vs 1.16 (xPW) and 1.34 vs 1.28 (sPW), and gCNR, 0.82 vs 0.77 (xPW) and 0.93 vs 0.89 (sPW).

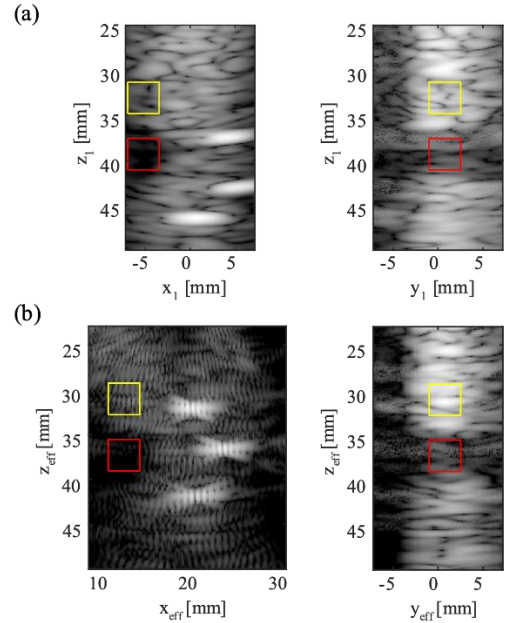

**Fig. 4.** Simulated *xz* and *yz*-plane B-mode images using the sPW sequence obtained for (a) the 1×512Tx/Rx (dynamic range 50 dB), and (b) the 2×256Tx/Rx CoMTUS (dynamic range 60 dB) configurations. The red squares represent the anechoic regions and the yellow squares the signal regions used for calculating the contrast metrics.

TABLE I
SIMULATION IMAGE QUALITY METRICS

| | Reference 1×512Tx/Rx (xPW) | CoMTUS 2×256Tx/Rx (xPW) | Reference 1×512Tx/Rx (sPW) | CoMTUS 2×256Tx/Rx (sPW) |
|---|---|---|---|---|
| CR [dB] | -13.8 | -16.7 | -17.1 | -22.1 |
| CNR | 1.16 | 1.23 | 1.28 | 1.34 |
| gCNR | 0.77 | 0.82 | 0.89 | 0.93 |
| FWHMx [mm] | 3.60 | 0.38 | 3.51 | 0.38 |
| FWHMy [mm] | 3.57 | 3.68 | 3.53 | 3.60 |
| FWHMz [mm] | 0.80 | 0.80 | 0.80 | 0.80 |

For the experimental data, CoMTUS images were beamformed after optimizing the relative position of the arrays and the local speed of sound of the medium using the backscattered echoes from the three steel spheres. The optimization algorithm was initiated with the theoretical values used in simulations and to mount the arrays on the 3-D printed holder (equation (1)). After the algorithm convergence, the local speed of sound was estimated as 1494 m/s, and the optimal values for the translation and the rotation angles that define the relative position of the probes were $\boldsymbol{T} = [34.1, -0.5, 19.9]^T$ mm, $\{\theta_x, \theta_y, \theta_z\} = \{-1.5º, -50.3º, -3.4º\}$. Note that the optimized beamforming parameters differ from the theoretical values used in the simulations. In particular, during the experiments, the arrays were not perfectly aligned with any axes due to the tolerances of the 3-D printer. The optimum CoMTUS images beamformed with this set of parameters are shown in Fig. 5-c. Compared with the 1×512Tx/Rx configuration, the experimental CoMTUS images show an extended FOV and a change in speckle texture in the *xz*-plane. As shown in Table II, all image quality metrics, but CR, improved in the experimental CoMTUS images. With the sPW scan sequence, the CNR increased from 0.90 to 1.21, and the gCNR from 0.76 to 0.84, while CR is approximately 1.5 dB lower. As observed in simulations, the sPW scan sequence resulted in higher performance than the xPW one.

Fig. 5 shows a comparison between the single transducer 1×512Tx/Rx (Fig.5-a) and CoMTUS images beamformed using the initial estimate of the parameters (Fig.5-b) and their optimum values (Fig.5-c). It is clearly shown that image quality in the *xz*-plane is degraded when the initial estimate of the parameters is used. There is an evident resolution loss in the *x*-direction and the targets are no longer successfully beamformed (Fig. 5-b) compared to the optimized solution (Fig. 5-c). When the theoretical beamforming parameters estimated from image registration are used, the different echoes backscattered from the targets do not properly align, creating interference when forming a coherent addition of signals. However, after optimizing the beamforming parameters in CoMTUS, all echoes substantially align and can be coherently added together in the same PSF shape predicted by simulations. Consequently, the image quality metrics improved after optimization. In the optimum case, the lesion is identifiable from the background signal with a higher contrast (-17.4dB vs -16.2dB) and gCNR (0.84 vs 0.80). There is an improvement in the speckle texture too.

To demonstrate the feasibility of free-hand operation, an operator freely moved the two probes in the 3-D printed holder changing the relative position of the probes, while raw channel data were continuously acquired for 2 seconds. Images were reconstructed offline considering the optimal beamforming parameters for each transmission, where each optimization was initialized with the optimum value of the previous transmission event. The resulting video, showing the sequence of successfully optimized frames, can be found as supplementary material.

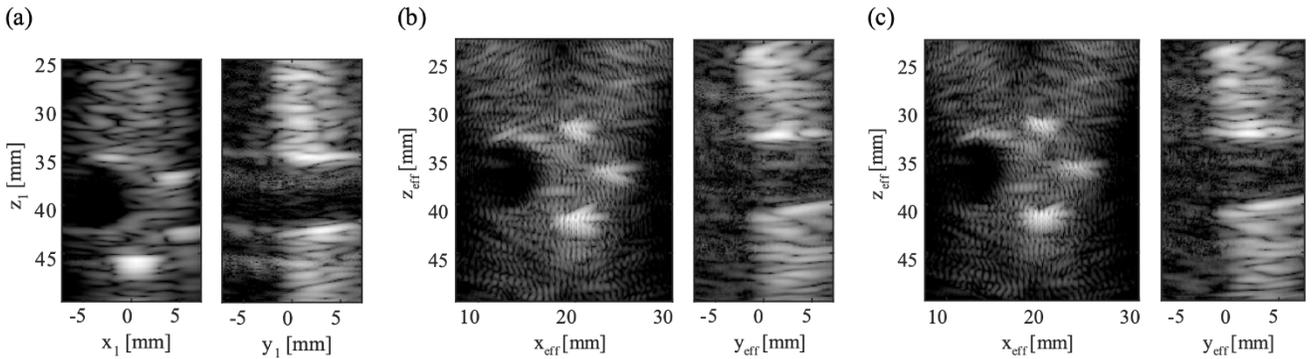

**Fig. 5.** Experimental *xz* and *yz*-plane B-mode images acquired on the tissue-mimicking phantom using the sPW sequence. (a) Images acquired with the single probe 1×512Tx/Rx configuration. (b) CoMTUS images beamformed with the theoretical values used in simulations ($\boldsymbol{T} = [34, 0, 14.4]^T, \{\theta_x, \theta_y, \theta_z\} = \{0º, -50º, 0º\}$). (c) CoMTUS images beamformed with the optimum parameters ($\boldsymbol{T} = [34.1, -0.5, 19.9]^T, \{\theta_x, \theta_y, \theta_z\} = \{-1.5º, -50.3º, -3.4º\}$). Dynamic range 50 dB.

TABLE II
EXPERIMENTAL IMAGE QUALITY METRICS

|  | Reference 1×512Tx/Rx (xPW) | CoMTUS 2×256Tx/Rx (xPW) | Reference 1×512Tx/Rx (sPW) | CoMTUS 2×256Tx/Rx (sPW) |
|---|---|---|---|---|
| **CR [dB]** | -15.1 | -13.8 | -18.9 | -17.4 |
| **CNR [-]** | 0.89 | 1.12 | 0.90 | 1.21 |
| **gCNR [-]** | 0.67 | 0.75 | 0.76 | 0.84 |
| **FWHMx [mm]** | 2.80 | 0.26 | 2.78 | 0.26 |
| **FWHMy [mm]** | 2.71 | 2.78 | 2.70 | 2.78 |
| **FWHMz [mm]** | 1.14 | 1.12 | 1.14 | 1.11 |

The improvements in resolution due to the extended aperture created by CoMTUS in the *x*-direction are evident in the PSF. Fig. 6 shows the PSF of the deepest target for simulations (Fig. 6 a-b) and experiments (Fig. 6 c-d) acquired with both probe configurations. There is a good agreement on the shape of the PSF between simulations and experiments, even if the resolution in *x* and *y*-directions are better in the experiments, while the simulations predict a better axial resolution (*z*-direction), as shown in Tables I and II. In comparison with the 1×512Tx/Rx configuration, CoMTUS improves resolution up to 10 times in the *x*-direction, along which the resulting effective aperture is extended. For both simulations and experiments, the scan sequences and, therefore, the number of compounding angles, do not seem to significantly affect the spatial resolution, even if a neglectable improvement could be seen for the 1×512Tx/Rx configuration. Differences in *y* and *z*-directions between simulations and experiments may be due to the different spatial configurations of the arrays. Results suggest that, when the aperture is extended in a particular direction, the resolution is mostly dictated by the aperture size and not by the scan sequence.

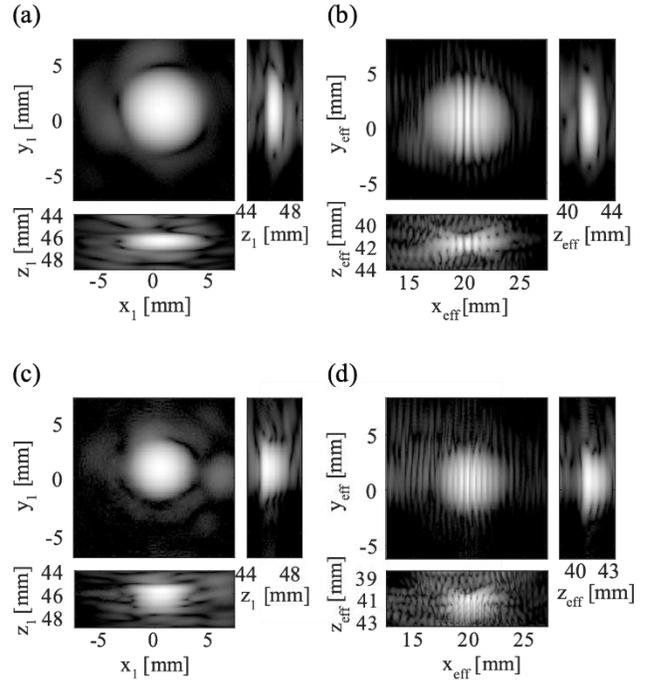

**Fig. 6.** (a-b) Simulated and (c-d) experimental PSF of the deepest target acquired using the sPW sequence, with 1×512Tx/Rx (a-c) and CoMTUS (b-d) configurations, respectively. Dynamic range: 50 dB.

## IV. DISCUSSION

In this study, the first demonstration of 3-D CoMTUS imaging is presented. The method was previously experimentally demonstrated only with linear arrays constrained to the same 2-D imaging plane [11]. In the present study, the experimental implementation of 3-D CoMTUS was possible by using two 512-element probe prototypes. As shown in Fig. 1, the prototypes were hardwired to allow the control of either two independent 256-element sparse arrays (2×256Tx/Rx configuration) or a single 512-element dense array (1×512Tx/Rx configuration) by using two synchronized ULA-OP 256 scanners [32]. The spiral sparse solution allows the channel count to be kept within feasible bounds and guarantees uniform performance over a wide range of steering angles [29]. In addition, the use of sparse arrays does not require channel to element multiplexing thus simplifying and keeping short the scan sequence, which is of uttermost importance for the CoMTUS scan sequence where the different transducers take turns to transmit PWs. The imaging performance of the method was investigated by both simulations and experiments. Two different scan sequences with a maximum Tx angle of 5° and a total of 9 and 22 PWs steered to form a cross-shape (xPW) and a 2-D spiral (sPW), respectively, were implemented.

Overall, CoMTUS shows narrower main lobe and higher contrast-to-noise-ratio than the single dense array, which results in an increased dynamic range and better target detectability. There were qualitative agreements between simulated and experimental PSFs and speckle textures (Fig. 4 and 5). CoMTUS contrast metrics on simulated data outperform experimental results, while resolutions in *x* and *y*-directions are better in the experimental case. These discrepancies are due in part to the simplified simulations. The experimentally measured worse axial resolution (*z*-direction) could be ascribed to the ring-down artifacts generated by the metallic spheres [40] that are present in the experimental images but cannot be simulated by Field II (Fig. 6). In addition, some other effects such as reverberations and direct transmissions between arrays, which cause image artifacts, could not be simulated by Field II. A simulation approach based on wave-propagation could better predict those effects [41][46], but would have excessively increased the model complexity and the simulation time.

CoMTUS performance was validated versus a single dense spiral array system using the same total number of active elements (1×512Tx/Rx). Both simulated and experimental results show that CoMTUS leads to images with extended FOV and significant improved resolution in the direction where both arrays are aligned, higher CNR and gCNR. The extended FOV and the improvements in lateral resolution (*x*-axis) were expected, as the spatial configuration of the arrays in CoMTUS results in an effective extended aperture in the *x*-direction (Fig. 2). In addition, results show improved CoMTUS imaging sensitivity with higher CNR and gCNR than the dense array system (Table II). Also, since the receive aperture is larger than in a single array system, CoMTUS could help to improve the low SNR of sparse arrays. In experiments, opposite to the simulations, there was a slight loos of CR (of 1.5 dB) for CoMTUS images compared to the 1×512Tx/Rx images, which may be mostly due to reverberation and cross-talk generated by direct transmissions between arrays.

Unlike resolution, improvements in CR and CNR were difficult to predict. Previous studies in 2-D imaging showed that the created discontinuous effective aperture degrades CR and CNR when the gap in the aperture is bigger than a few centimeters [12]. In the present configuration, the gap between the arrays was approximately 38 mm, already large enough to affect CR and CNR [12]. However, very preliminary simulated results showed improvements in CR and CNR in 3-D imaging for a different probe configuration and scan sequence [42]. This suggests that non-periodic apertures have the potential to benefit CoMTUS and there may be an opportunity to distribute the number of resources available to reduce grating-lobe artifacts and improve image quality [43]. The interplay between the relative location of 2-D arrays, the effect of apodization, and the resulting resolutions and contrast will be the focus of future studies.

The quality of CoMTUS images also depends on the accuracy of the beamforming parameters, which include the transducers' location and the local speed of sound. Results show that an accurate estimation is essential to guarantee successful performance and that a calibration based only on the use of probe holders will not be good enough for an outstanding image (Fig. 5). The required accuracy was achieved by maximizing the coherence of received backscattered echoes arising from the three steel targets in the medium, as previously described in [11]. The initialization of the algorithm is comparable with the 2-D imaging case, but with a 3-D image-based registration. The subsequent running times for the optimization are slightly higher than for 2-D imaging data because of the larger number of channels used. As previously observed, optimization between consecutive acquisitions, where each optimization is initialized with the output from the previous one, converges significantly faster (free-hand demonstration). However, how this method will perform in more complex scenarios is still unclear and further studies are needed to predict the performance for in vivo imaging. Preliminary in vivo results calibrated on anatomical features are promising [13]. In addition, large aperture studies with an external tracker showed a gain for deep imaging [44] and it is expected that CoMTUS should only improve on this. Previous studies with two curved array transducers show that multi-perspective images calibrated using image registration still can improve the reconstruction of particularly challenging applications such as the aortic wall and provide significant benefits for strain imaging [9]. Bottenus *et al.* [10] quantified the accuracy required to coherently merge information from different probe locations in 2-D imaging. For the particular case of multiple 2-D arrays, further investigations are needed. Future work will assess the effect of calibration errors on each imaging dimension.

Although initial in vivo results are promising, there are still many challenges to address in order to achieve a full practical deployment of 3-D CoMTUS. Indeed, the actual translation of any multi-transducer 3-D imaging modality to the clinic is hindered by intrinsic limitations, i.e.: the limited FOV due to the small overlapping volume; the need for multiple probes, which are not easy to place, handle and maneuver, even with a holder; and the lack of real-time feedback, due to the current complexity of the scan sequence and the beamforming algorithms. Therefore, in the future, different Tx beam profiles, such as diverging waves [45], could be investigated to increase the FOV. At the same time, wider FOV might help the probe positioning as it would be easier to find overlap between insonated regions, thus making easier the determination of the relative probe positions by the optimization algorithm.

On the other hand, the FOV could be extended by the use of more than two matrix arrays, which will also improve the resolution in all directions and potentially solve the anisotropic resolution problem in multi-transducer volumetric imaging. In addition, although it could seem counterintuitive, the use of several distributed probes might also solve the issues of placing, handling and maneuvering the probes. Indeed, the FOV could be big enough to provide a comprehensive and panoramic 3-D view of the region of interest at any time, without the need of moving the probe. Therefore, the probes will not need to be handheld but could be mounted on a fixed, but adaptable, support and directly placed over the region of interest.

Finally, for clinical exploitation of CoMTUS, substantial developments should be done to reduce its computational load and the amount of data to process, which are significantly large and increase with the number of transducers. Parallel-beamforming, graphical processing unit (GPU)-based platforms, and high-speed buses will be extremely important in the future to achieve real-time capabilities [46][47], but also algorithms for data compression and reduction will be needed [48][49].

V. CONCLUSION

In this study, the feasibility of CoMTUS for 3-D imaging is demonstrated for the first time using a pair of 2-D sparse spiral arrays. The use of the sparse arrays allows the channel count and the amount of data to be kept within feasible bounds, which is extremely important for the actual implementation of a multi-array system. CoMTUS images were acquired on a tissue-mimicking phantom and compared with the images acquired by a dense array with the same number of active elements, i.e., a total of 512 elements. The performance of the system was evaluated by both simulations and experiments. In comparison to a single dense spiral array system, CoMTUS leads to improved images with an extended FOV, improved resolution in the direction where both arrays are aligned, and higher CNR and gCNR.

ACKNOWLEDGMENT

The authors would like to thank the members of the Microelectronics systems design laboratory (MSDLab) at the Department of Information Engineering of the University of Florence for their technical support during the experimental data collection.